\documentstyle[aps,multicol,epsfig]{revtex}
\newcommand{\nc}{\newcommand}
\nc{\be}{\begin{equation}}
\nc{\ee}{\end{equation}}
\nc{\bea}{\begin{eqnarray}}
\nc{\eea}{\end{eqnarray}}
\nc{\bean}{\begin{eqnarray*}}
\nc{\eean}{\end{eqnarray*}}
\nc{\mb}{\mbox}
\nc{\rnc}{\renewcommand}
\nc{\r}{\mb{\boldmath$r$}}
\nc{\x}{\mb{\boldmath$x$}}
\nc{\A}{\mb{\boldmath$A$}}
\nc{\sa}{\mb{\boldmath$a$}}
\nc{\nab}{\nabla}
\nc{\X}{\sf x}

\begin{document}
\draft

\def\del{\partial}

\title{ Pairing symmetry transitions in the even-denominator FQHE system
}
\author
{Kentaro Nomura\cite{email} and Daijiro Yoshioka
 }

\vspace{5mm}

\address{
Department of Basic Science, University of Tokyo, 3-8-1 Komaba, Tokyo 153-8902\\
}

\vspace{4mm}

\date{\today}
\maketitle

\vspace{0.5cm}

\begin{abstract} 
 Transitions from a paired quantum Hall state to another
quantum Hall state in bilayer systems are discussed 
in the framework of the edge theory.
Starting from the edge theory for the Haldane-Rezayi state,
it is shown that the charging effect of a bilayer system
which breaks the SU(2) symmetry of the pseudo-spin
shifts the central charge and the conformal dimensions 
of the fermionic fields which
describe the pseudo-spin sector in the edge theory. 
This corresponds to the transition from Haldane-Rezayi
 state to Halperin's
331 state, or singlet d-wave to triplet p-wave ABM type paired state
in the composite fermion picture.
 Considering interlayer tunneling, 
the tunneling rate-capacitance
 phase diagram for the 
$\nu=5/2$ paired bilayer system is discussed.
\end{abstract}

\vspace{0.5cm}



\begin{multicols}{2}
\narrowtext

   A bilayer quantum Hall system can be mapped to an 
equivalent spin-1/2
system by assigning $\uparrow$($\downarrow$) pseudo spins
to electrons in the upper (lower) layer, \cite{eis2}
where the 
actual electron spins are assumed to be polarized.
At finite layer separation, the Hartree part of the
electron-electron repulsive interaction produces
a local capacitive charging energy that is minimized
when the two layers have equal electron density.
Especially for the $\nu=1/$(odd integer) quantum Hall
ferromagnet, the expectation value of the z-component of the
 pseudo spin vanishes and the pseudo spin system has 
an easy-plane anisotropy that gives the itinerant 
ferromagnet with an XY symmetry in the absence of interlayer
tunneling.
The characteristic parameters are the distance
 between the layers, the interlayer
tunneling rate, and the electric capacitance etc.
The grand state is characterized by these parameters.
The simplest Abelian quantum Hall state for the bilayer
system is described by the two-component generalization 
of  the Laughlin wave function first introduced by 
Halperin as \cite{331}
\bea
 \Psi_{mmn}= \prod_{i<j}
(z_i^{\uparrow}-z_j^{\uparrow})^m
\prod_{i<j}(z_i^{\downarrow}-z_j^{\downarrow})^{m}
\prod_{i,j}(z_i^{\uparrow}-z_j^{\downarrow})^n, \nonumber \\
\eea
where we abbreviated the factor 
${\rm e}^{-\frac{1}{4}\sum(|z_i^{\uparrow}|^2+|z_i^{\downarrow}|^2)}$.
In the case of $\nu=1/2$, $(m,m,n)=(3,3,1)$, and the wave
 function is rewritten as 
\bea
  \Psi_{331}={\rm {det}}
\left( \frac{1}{z_i^{\uparrow}-z_j^{\downarrow}} \right)
  \Psi_{222}.
\eea
This state is called $331$-state.
 
 Haldane and Rezayi have proposed a two-component
singlet wave function to explain the quantum Hall effect at
$\nu=5/2$. \cite{hr,ymg}
 This state is an exact grand state of the 
hollow-core model and is given by
\bea
 \Psi_{\rm HR}
={\rm {det}}
\left( \frac{1}{(z_i^{\uparrow}-z_j^{\downarrow})^2}
\right)
  \Psi_{222},
\eea
which is called Haldane-Rezayi(HR) state.

In above two states the electron spins are unpolarized.
Moore and Read proposed spin polarized state which is 
described by the trial function; \cite{mr}
\bea
 \Psi_{\rm Pf} = {\rm Pf}\left( \frac{1}{z_i-z_j} \right)
   \prod_{i<j}(z_i-z_j)^2,
\label{Pf}
\eea
This state is called Pfaffian state.  ${\rm Pf}(M_{ij})$ means
${\it A}\prod_i M_{2i-1,2i}$.
Greiter et al. studied  the Pfaffian state as a paired state of 
the composite fermions. \cite{gww,jain}
  Ho considered interesting 
connections between the  internal order in these types of wave 
functions and analogous order in superfluid $^3$He. \cite{ho}
The order parameter symmetry is triplet p-wave with ABM and A$_1$
type in $^3$He for 331 and Pfaffian state, respectively.
Similarly HR state can be regarded as d-wave paired state.\cite{rg}

  The wave functions for the grand state and the excited states
 in these paired states are interpreted as the correlation functions of
 proper conformal field theories(CFT). In the case of the Pfaffian, 331, and 
HR states, corresponding theories have a central charge
$c= 1+1/2, 1+1,1-2$.  \cite{mr,ww,mir,lw}
These conformal theories also
 describe the edge excitations on 
the boundaries of the sample. 
 In the even-denominator quantum Hall regime such as $\nu=1/2$ and $5/2$,
the grand state was used to be considered as one of these 
three states.
Recently, Read and Green argued that the HR state is right 
at bulk transition point between weak and strong pairing
d-wave sates.\cite{rg} 
In this paper we discussed another direction of the developments
which is the grand state of the bilayer systems.
Let us consider the charging effect and interlayer tunneling
which are described in terms of the pseudo spin as
\bea
 H_{\rm U} =
\epsilon_{\rm c}
\int {\rm d}^2\x \left( 
 S^z(\x)
\right)^2,
\label{Hu}
\eea
and  
\bea
 H_{\rm T} =  -\Delta_{SAS} \int {\rm d}^2\x S^x(\x),
\label{Ht}
\eea
respectively.
Latter is regarded as Zeeman term for pseudo-spin.
It is well known that  introducing a Zeeman term
the pseudo-spins align to the direction of the (pseudo)
magnetic field, and pseudo-spin polarization transition 
occurs. This process corresponds to the 331-Pfaffian transition
which is argued in  Ref 8).
 On the other hand, we can expect that if the introduced anisotropy
is XY (i.e. easy-plane) type, the singlet-triplet transition
might appear.

 In this paper we describe the latter transition which  
occurrs between the HR and 331 due to the charging effect
in the bilayer system with hollow-core interaction which
is perturbed by layer separation
in the framework of the edge theory. 
Our results predict that the 331 state is favored for
the charging term.
This is consistent with a numerical studies \cite{ymg2,scarola} 
for realistic interaction.
 
 The edge theory is first established 
by X.G. Wen \cite{wen1,wen2} and the connection with the bulk wave function
is indicated by Moore and Read.\cite{mr}

 First we review this connection for the $\nu=1/q$ Laughlin state.
 The grand state wave function is written as 
the correlation function of the $c=1$ CFT.
\bea
 \Psi_{\rm Ln}= <\prod_{i=1}^{N_e} \psi_e(z_i)
\Phi_{\Delta}>,
\eea
where $ \psi_e(z)=:{\rm e}^{{\rm i}{\sqrt q}\phi(z)  }:$
is the electron field operator,
$\Phi_{\Delta}=:{\rm e}^{-{\rm i} 
\rho_0\int d^2z {\sqrt q} \phi(z)  }:$
is the primary field which has the conformal dimension
$\Delta=qN_e/2$, and we ignore the singular phase which is derived from
the right hand side of eq.(7). The
$c=1$ CFT is represented by the action of the free bosonic field
\bea
 S= \int {\rm d}^2x \frac{1}{2\pi} \partial \phi {\overline 
{\partial}} \phi,
\eea
where $z$ is the complex coordinate,
 the correlation exponent 
of the theory is given as $K=\nu$.
  The edge excitations are connected with the 
angular momentum excitations of the bulk.
For a disc geometry, the Hamiltonian for edge modes is given as
\bea
 H = \frac{2\pi v}{L} L_0,
\eea
where 
$ L_0 = \frac{1}{2} J_0^2 + \sum_{n>1}^{\infty}
J_nJ_{-n}$
is the holomorphic part of the zero-component of the Virasoro algebra
which  can be regarded  as the
z-component of the  total orbital angular momentum,
$J$ is the Kac-Moody current, 
and we ignore
the correction term which is proportional to the 
central charge.

Now we come back to the present bilayer case.
 The edge theory of the paired quantum Hall state
has an internal degree of freedom which corresponds
to the pseudo spin in the bilayer systems.
The additional prefactor to Laughlin-Jastrow type factor 
of the three paired state
wave functions in eqs.(2)-(4) arise from these pseudo spin mode.

  The edge theory for the 331 state is described
by the Hamiltonian
\bea
 H_{331}= \frac{1}{4\pi} \int {\rm d}x \ v_{IJ} 
 :\partial_x \phi_I
\partial_x \phi_J :,
\eea
where $v_{IJ}$ is a symmetric matrix which
depends on the confining potential and interchannel 
interactions at the edge.\cite{mr,wen2}
 The commutation relations of the bosonic fields are
 $[\phi_I(x),\phi_J(x')]= {\rm i}\pi K_{IJ} {\rm sgn}(x-x')$,
where $K$ is the K-matrix given as \cite{wz}
\bea
 K = \left(
  \begin{array}{@{\,}cc@{\,}}
   3 & 1 \\
   1 & 3
 \end{array}
 \right) .
\eea
 In the charge-pseudo spin basis, 
$ \phi_c = \frac{1}{\sqrt {\nu}}
(\phi_{\uparrow} + \phi_{\downarrow}),\ \
 \phi_s = 
\phi_{\uparrow} - \phi_{\downarrow}$,
the two components can be separated as
\bea
 H_{331}=\frac{1}{4\pi}\int {\rm d}x \left[  v_c
 : \left( \partial_x \phi_c \right)^2 :+
v_s  :\left( \partial_x \phi_s\right)^2:
 \right].
\eea
Both charge and spin sectors are also written as
the free bosonic form which have the correlation exponent
$K_c=\nu, K_s=1$, respectively.
The electron operator is given by
$ \psi_{e\uparrow}= {\rm e}^{\frac{\rm i}{\sqrt {\nu}}\phi_c
+{\rm i}\phi_s},\
 \psi_{e\downarrow}= {\rm e}^{\frac{\rm i}{\sqrt {\nu}}\phi_c
-{\rm i}\phi_s}$.
 Because the case $K_s=1$ is equivalent to
the free fermion theory, the spin sector Hamiltonian is also
described by
\bea
 H^s_{331} = -{\rm i}v \int {\rm d}x \  \psi^{\dag} \partial_x \psi 
= \frac{2\pi v}{L} L_0,
\eea
where the fermionic field $\psi$ is related to the 
bosonic field by 
$\psi={\rm e}^{{\rm i} \phi_s}$, and
 $L_0 = \frac{1}{2} (J_0^z)^2 + \sum_{n>1}^{\infty}
J^z_nJ^z_{-n}$,  $J^z ={\rm i}\partial \phi_s$ is the z-component
of the pseudo-spin current density on the edge.

As we mentioned above, 
the spin sector corresponds to the degrees of 
the freedom of the 
pairing.
 Actually the equality
\bea
{\rm  det}\left( \frac{1}{z_i^{\uparrow}-z_j^{\downarrow}
} \right) 
=<\prod_{i,j} \psi^{\dag}(z_i^{\uparrow})
\psi(z_j^{\downarrow}) >
\eea
can be derived.

In the HR state,
the electron operator is written as 
$ \psi_{e\sigma}=\partial\theta_{\sigma}
 e^{{\rm i} {\sqrt {2}}\phi_c}$,
The determinant part of the wave function is reconstructed 
by the conformal field theory of the fermionic ghost field as
\bea
 {\rm det}\left( \frac{1}{(z_i^{\uparrow}-z_j^{\downarrow})^2} \right)
= <\prod_{i,j}  \partial \theta_{\uparrow}
(z_i^{\uparrow}) 
\partial \theta_{\downarrow}(z_i^{\downarrow}) 
>,\nonumber 
\eea
that corresponds to the spin sector of the edge excitation.
On the other hand, the charge sector is  same as the theory 
for the 331 state.
 The spin sector of the edge excitations is described by  
\bea
 \theta_{\uparrow} &=& {\sqrt {\frac{2\pi}{L} }}\sum_{k>0}
\frac{1}{\sqrt {k}}(c_{k\uparrow} {\rm e}^{{\rm i} kx}
+ c_{k\downarrow} {\rm e}^{- {\rm i} kx})\nonumber \\
 \theta_{\downarrow} &=& {\sqrt {\frac{2\pi}{L} }}\sum_{k>0}
\frac{1}{\sqrt {k}}(c_{k\uparrow} {\rm e}^{ {\rm i} kx}
- c_{k\downarrow} {\rm e}^{- {\rm i} kx}).
\eea
The Hamiltonian is given as \cite{mir,lw}
\bea
 H^s_{\rm HR} &=& \int {\rm d}x \frac{v_s}{4\pi}:(\partial_x\theta_{\uparrow}
\partial_x\theta_{\downarrow}   +  
  \partial_x\theta_{\uparrow}^{\dag}
\partial_x\theta_{\downarrow}^{\dag}
):\nonumber \\
 &=& \sum_{k>0} v_s k \left(  
c_{k\uparrow}^{\dag} c_{k\uparrow}  +
c_{k\downarrow}^{\dag} c_{k\downarrow} 
\right) ,
\label{Hhr}
\eea 
which has the global SU(2) symmetry with the total spin generator
\bea
 S_z &=& \frac{1}{2}\sum_{k>0}(
c_{k\uparrow}^{\dag} c_{k\uparrow}  -
c_{k\downarrow}^{\dag} c_{k\downarrow} 
),   \nonumber \\
 S_{+} &=& \sum_{k>0} c_{k\uparrow}^{\dag} c_{k\downarrow},\ \ \ 
S_{-} = \sum_{k>0} c_{k\downarrow}^{\dag} c_{k\uparrow}.
\label{spin}
\eea
To go on our argument further,
we convert to the
$(\xi, \eta)-$representation:
\bea
 S^s_{\rm HR}=\int  {\rm d}^2x  \frac{1}{2\pi} \left(  
\eta {\overline {\partial}} \xi + 
{\overline {\eta}}\partial
{\overline {\xi}}
\right),
\label{action}
\eea
where the field $\eta=\partial\theta_{\uparrow}$ and
$\xi=\theta_{\downarrow}$ are fermionic field which 
satisfy
\bea
 <\xi(z) \eta(\omega)> = \frac{1}{z-\omega}.
\eea
The conformal dimensions of the 
$\xi$ and $\eta$ are 0, 1, respectively.
The energy-momentum tensor is derived from (\ref{action}) as
\bea
 T(z)&=& :\partial \xi \eta(z) : \nonumber \\
  &=& -\frac{1}{2} :(\eta\partial\xi-
\partial\eta\xi)(z):
+ \frac{1}{2} :(\partial \xi \eta + \xi \partial \eta)(z) :
\nonumber \\
&=& \frac{1}{2}J^z(z)^2 + \frac{1}{2}\partial J^z(z),
\eea
where $J^z(z)=:\xi \eta(z):$.
Calculating the operator product expansion;
\bea
 T(\omega)\eta(z) &=& \frac{1}{(\omega -z)^2}\eta(z)
+\frac{\partial \eta(z)}{\omega -z}
+ \cdots,   \\
 T(\omega)\xi(z) &=& \frac{0}{(\omega-z)^2}\xi(z)
+\frac{\partial \xi(z)}{\omega -z}
+ \cdots,   \\
T(\omega)T(z)&=& \frac{-1}{(\omega- z)^4} +
\frac{2T(z)}{(\omega- z)^2} + \cdots,
\eea
we can check the conformal dimension of the 
field $\eta$ and $\xi$ and the central charge $c=-2$ directly.
The zero-component of the energy-momentum tensor is 
\bea
 L_0^{c=-2}
 &=& \frac{1}{2} J^z_0(J^z_0-1) + \sum_{n>0}^{\infty}
J^z_n J^z_{-n}.
\eea
Then the 
spin sector of the edge Hamiltonian for HR-state eq.(\ref{Hhr}) is 
also written as
\bea
 H_{\rm HR}^{s}= \frac{2\pi v}{L} L_0^{c=-2}.
\eea

Now we consider the charging effect.
On the edge, the
charging Hamiltonian (\ref{Hu})
is rewritten as 
\bea
 H_{\rm U} = \int {\rm d}x \ gJ^z(x)^2 
\eea
where a constant $g$ is 
 proportional to $\epsilon_{\rm c}$ in (\ref{Hu}).

The total Hamiltonian is given by
\bea
 H  &=&H^{c} + H_{\rm HR}^{s} + H_{\rm U} \nonumber \\
    &=& H^c + \frac{2\pi v_s}{L}(
\frac{1}{2}(J_0^z)^2 + \sum_{n>0}^{\infty}J^z_n J^z_{-n}
-\alpha J^z_0).
\label{totalH}
\eea 
where $v_s=v+gL/2\pi$ is the renormalized velocity,
and $\alpha=v/(v+gL/2\pi)$.
For a large $g$, 
 the last term of the eq.(\ref{totalH}) can be ignored and 
the spin sector is modified as below.
The energy-momentum tensor which derive the Hamiltonian
eq.(\ref{totalH}) with $\alpha=0$ is 
\bea
 T_{\rm eff}(z) 
=
\frac{1}{2} \left(J^z(z)  \right)^2 \nonumber
=
- \frac{1}{2} :(\eta \partial \xi
- \partial \eta  \xi):
\eea
This mechanism is called spectral flow which originally
discussed in Ref\cite{gfn,gl,ino1,ino2,cap}.
 The conformal dimensions of the fermion fields 
become
$\Delta_{\xi}=\Delta_{\eta}=1/2$.
Using Wick's theorem we can check it in the operator 
product expansion:
\bea
 T_{\rm eff}(\omega)\eta(z)
&=& \frac{1/2}{(\omega -z)^2}\eta(z) +
 \frac{\partial \eta(z)}{\omega -z} + \cdots,   \\
 T_{\rm eff}(\omega)\xi(z)
&=& \frac{1/2}{(\omega -z)^2}\xi(z) +
 \frac{\partial \xi(z)}{\omega -z} + \cdots.
\eea
Not only the conformal dimensions of the fermion fields
but  the central charge of the spin sector
is also shifted from $c=-2$ to unity.
 So we can rewrite the fermionic field as
\bea
 \psi &\equiv & \eta = 
{\sqrt {\frac{2\pi}{L} }}\sum
f_k {\rm e}^{ {\rm i} kx}
\nonumber \\
 \psi^{\dag} &\equiv& \xi =
{\sqrt {\frac{2\pi}{L} }}\sum
f_k^{\dag} {\rm e}^{- {\rm i} kx},
\eea
and the Hamiltonian as
\bea
 H^s_{\rm eff} &=& 
\int \frac{{\rm d}x}{2\pi} \ \psi^{\dag}\left(
-{\rm i}v_s \frac{\partial}{\partial x}
\right)\psi 
\nonumber \\   &=&
\sum_k v_s k\  f_k^{\dag} f_k.
\label{Heff}
\eea
In eq.(\ref{Heff}), the global SU(2) symmetry represented
 by eq.(\ref{spin})
is reduced to U(1) XY symmetry.
This modified theory can be reconciled with the spin sector
 for the 331 state.
Therefore due to the charging effect,
 the HR-331 transition occurs.

Next we discuss the effect of interlayer tunneling (\ref{Ht}) on the edge
that  causes the pseudo-spin polarization
to $x$ direction. 
 So the Hamiltonian (\ref{Hhr}) become
that for the single component 
\bea
 H^s_{\rm Pf} &=& \sum_{k>0} v_s k \ c_{k+}^{\dag} c_{k+} \nonumber \\
   &=& \int \frac{{\rm d}x}{2\pi} \  \chi \left( 
-{\rm i}v_s \frac{\partial}{\partial x}
  \right) \chi,
\eea
 where 
$c_{k\pm}= \frac{1}{\sqrt {2}}(c_{k\uparrow} \pm c_{k\downarrow})$
and
\bea
\chi= {\sqrt {\frac{2\pi}{L} }}\sum_{k>0}
(c_{k+} {\rm e}^{ {\rm i} kx}  
+ c_{k+}^{\dag} {\rm e}^{-{\rm i} kx}
).
\eea
The spin sector $\chi$ have the correlation function
\bea
{\rm  Pf}\left( \frac{1}{z_i-z_j} \right) 
=< \chi(z_1) \cdots
\chi(z_N) > ,
\eea
that contributes to the bulk wave function (\ref{Pf}).
 So these process is regarded as
the polarization transition between the HR and Pfaffian 
state. A edge theory discription of the 331-Pfaffian transition
 is discussed in Ref. [\ref{rg}].

Finally we consider the tunneling amplitude versus
layer separation $d$ phase diagram.
We regard these two parameter as independent.
When the interlayer tunneling rate is large.
the system has effectively one component. (i.e.
pseudo spin is polarized to the easy-plane.)
This  is the Pfaffian state.
Contrary, in the regime where the tunneling amplitude
is small, the 331 and HR state are expected to appear.
According to above analysis, the HR-331 transition occurs as 
 the charging effect is increased.
These statements are summarized as Fig.1.

 In this paper we have described the pairing symmetry transition
in the bilayer even denominator quantum Hall states in the framework
of the conformal field theories.
Starting from the edge theory of the HR state
 the charging term shifts the   central charge from $c=-2$
to $c=1$ and the conformal dimensions of the fermion fields.
This is interpreted as the HR-331 transition.
In view of the composite fermion theory,
HR state is the d-wave paired state and 331 state is 
the p-wave ABM state. So the transition is from singlet
to triplet. Our result shows that the 331 state is favored 
in the bilayer systems with the finete layer separation.
In the context 
of our edge theory description, we
cannot know what kind of the transition this belongs.
The detailed investigation of
 the transitional or crossover points
 is left to future problems.

%
%
\section*{Acknowledgements}
We are grateful to A. Yamaguchi and K. Hamada 
for useful
discussions and thier collaborations.
We also acknowledge K. Ino for pointing out the mechanism of the
spectral flow and N. Read for helpful indication for our phase diagram. 
This work is supported by a Grant-in-Aid for Scientific Research (C) 10640301
from the Ministry of Education, Science, Sports, Culture and Technology.
%
%
%
%
\begin{figure}
\epsfxsize 70mm \epsffile{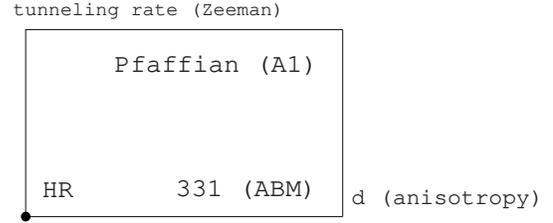}
\caption{
The proposed  phase diagram where $d$ is separation between
the layers.
The polarized state is  the Pfaffian state only. 
The unpolarized states are 
331 and HR state. The latter state exists as a critical point
(dot in the diagram).
}
\label{fig01}
\end{figure}

\end{multicols}


\begin{thebibliography}{99}
\bibitem[*]{email} nomura@sola.c.u-tokyo.ac.jp
\bibitem{eis2} J. P. Eisenstein in "{\it {
Perspectives in Quantum Hall Effects
}}" ed. by S. Das Sarma and A. Pinczuk, 1997, John Wiley 
and Sons, New York.

\bibitem{331} B. I. Halperin Helv.Phys. Acta. {\bf 56} (1983) 75.


\bibitem{hr} F. D. M. Haldane, R. Rezayi: Phys. Rev. Lett. {\bf 60} (1988) 956.

\bibitem{ymg} D. Yoshioka, A. H. MacDonald, and S. M. Girvin:
  Phys. Rev. B {\bf 38} (1988) 3636.
\bibitem{mr} G. Moore, N. Read: Nucl. Phys. B {\bf 360} (1991) 362.

\bibitem{gww} M. Greiter, X. G. Wen, and F. Wilczek: Nucl. Phys.  {\bf B374}
 (1992) 567.

\bibitem{jain} J. K. Jain: Phys. Rev. Lett. {\bf 63} (1989) 199.
 
\bibitem{ho} T. L. Ho: Phys. Rev. Lett. {\bf 75} (1995) 1186.

\bibitem{rg} N. Read and D. Green: Phys. Rev. B {\bf 61} (2000) 10267.
\label{rg}



\bibitem{ww} X.G. Wen and Y. S. Wu: Nucl. Phys. {\bf B419}
(1994) 455.

\bibitem{mir} M. Milovanovic and N. Read: Phys. Rev. B
{\bf 53} (1996) 13559.

\bibitem{lw} J. C. Lee and X.G. Wen: Nucl. Phys. {\bf B542}
(1999) 647. 

\bibitem{ymg2} D. Yoshioka, A. H. MacDonald and S. M. Girvin:
 Phys. Rev. B {\bf 39} (1989) 1932.
\label{ymg2}
\bibitem{scarola} V.W. Scarola and J.K. Jain:{\it {preprint}} cond-mat/0011511.

\bibitem{wen1} X. G. Wen: Phys. Rev. B {\bf 41} (1990) 12838.

\bibitem{wen2} X. G. Wen: Int. J. Mod. Phys. {\bf B6} (1992) 1711.

\bibitem{wz} X.G. Wen and A. Zee: Phys. Rev. B {\bf 46} (1992) 2290.

\bibitem{gfn} V. Gurarie, M. Flohr and Nayak: Nucl. Phys.{\rm B498} (1997) 513.

\bibitem{gl} S. Guruswamy and A. W. W. Ludwig: Nucl. Phys. {\rm B519} (1998) 661.

\bibitem{ino1} K. Ino: Nucl. Phys. {\rm B532} (1998) 782.

\bibitem{ino2} K. Ino: Phys. Rev. Lett. {\rm 82} (1999) 4903.

\bibitem{cap} A. Cappelli, L.S. Georgiev, and I. T. Todorov:
Commun. Math. Phys. {\bf 205} (1999) 657.



\label{cab}

\end{thebibliography}
\end{document}